\documentclass[useAMS,usenatbib]{mn2e}
\usepackage{aas_macros,graphicx,times,multirow,amsmath,bbold}
\usepackage[]{color}
\def\src{RX J1856}

\def \aap {A\&A}

\def \apjl {ApJL}

\title[Surface field structure in RX J1856]
{Probing the surface magnetic field structure in RX J1856.5-3754}
\author[S.~B.~Popov, R.~Taverna \& R.~Turolla]{S.~B.~Popov$^{1}$\thanks{E-mail:
\href{mailto:sergepolar@gmail.com}{sergepolar@gmail.com}}, R.~Taverna$^{2}$, R.~Turolla$^{2,\, 3}$
\smallskip\\
$^1$Sternberg Astronomical Institute,
Lomonosov Moscow State University, Moscow, Russia 119234\\
$^2$Department of Physics and Astronomy, University of Padova, via Marzolo 8, 35131 Padova, Italy\\
$^3$Mullard Space Science Laboratory, University College London, Holbury St. Mary, Surrey, RH5 6NT, UK\\
}

\date{Accepted \ldots. Received \ldots; in
original form \ldots} \pagerange{\pageref{firstpage}--\pageref{lastpage}} \pubyear{2013}

\def\LaTeX{L\kern-.36em\raise.3ex\hbox{a}\kern-.15em
    T\kern-.1667em\lower.7ex\hbox{E}\kern-.125emX}

\begin{document}

\label{firstpage}
\maketitle
\begin{abstract}
The evolution of magnetic field in isolated neutron stars is one
of the most important ingredients in the attempt to build a
unified description of these objects. A prediction of field
evolution models is the existence of an equilibrium configuration,
in which the Hall cascade vanishes. Recent calculations have
explored the field structure in this stage, called the Hall
attractor. We use X-ray data of near-by, cooling neutron stars to
probe this prediction, as these sources are surmised to be close
to or at Hall attractor phase. We show that the source RX
J1856.5-3754 might be closer to the attractor than other sources
of its class. Our modelling indicates that the properties of
surface thermal emission, assuming that the star is in the Hall
attractor, are in contradiction with the spectral data of RX
J1856.5-3754.
\end{abstract}

\section{Introduction}
\label{intro}

Despite isolated neutron stars (INSs) are mostly detected as
radiopulsars (PSRs), the existence of different manifestations of INSs, with
vastly diverse observational properties, emerged over the last two decades.
These include the soft $\gamma$-repeaters and the anomalous X-ray pulsars
\cite[SGRs and AXPs, aka the magnetar candidates; see e.g.][for
reviews]{mereg08,turolla15}, the central compact objects in supernova
remnants \cite[CCOs; e.g.][]{deluca08,ho13}, the thermally emitting X-ray
INSs \cite[also called the Magnificent seven, M7 hereafter;
e.g.][]{turolla09} and the rotating radio transients
\cite[RRaTs; e.g.][]{bspol13}.  The number of known objects in each class is
fairly limited ($\approx 10$--$100$) when compared to that of PSRs ($\approx
2000$ in the ATNF catalogue, \citealt{atnf})\footnote{The ATNF Pulsar Database,
{www.atnf.csiro.au/people/pulsar/psrcat/}.  }.  Still, estimates of the
birth rates indicate that some INS groups may outnumber PSRs \cite [the
thermally emitting X-ray INSs 
and the RRaTs in particular, see][]{popov06,kk08} and the total INS
birth rate turns out to exceed the commonly accepted value of the Galactic
supernova rate (\citealt{keane08}).  This has been taken as evidence that some evolutionary
links among the INS classes must exist and led to the quest for the
so-called ``grand unification'' of neutron stars, or GUNS for short
\cite[e.g.][]{kaspi10,igos14}.

The basic idea behind GUNS is that the observational status of an
INS at a given age depends on its properties at birth, ---
chiefly the mass, the initial spin
period, and the magnetic field, --- but not in a simple manner.
The successive thermal, rotational and
magnetic evolution is responsible for the outward appearance of the star,
which can pass through different stages, crossing from one class to another.
Thermal and magnetic evolution are inherently coupled, since the magnetic
diffusivity, which enters the induction equation for the $\mathbf B$-field,
strongly depends on temperature (and density).

Models of magneto-thermal evolution of NSs provide predictions which
can be tested using observational data. In particular, since
the crustal magnetic field dictates the surface temperature distribution and hence the properties of the thermal emission,
the observed spectra of thermally emitting INSs can be used to probe the topology of the magnetic field and its evolution.
In this paper we confront the predictions of the Hall attractor model
(\citealt{gc2013, gc2014_2}) with X-ray spectral data of the M7. In section \ref{hall} we
briefly summarize the properties of the Hall evolution and of the Hall attractor, while in section \ref{pop} we discuss which classes of INSs are best suited to this kind of investigation.
Our model for the surface emission is presented in section \ref{thermal} and results in section \ref{results}. Discussion follows in section \ref{disc}.


\section{Magnetic field evolution and the Hall attractor}
\label{hall}

In recent years, magnetic field evolution in INSs has been the focus of
several studies (see a review in \citealt{geppert2009}).   As noticed by
\cite{ponsgep07}, the internal magnetic field of INSs is likely supported by
two current systems, a short-lived one in the crust, with typical decay time
$\tau_\mathrm{decay}\approx 0.1$--$10$ Myr and a long-lasting one in the
core, $\tau_\mathrm{decay}\ga 10$--$100$ Myr \cite[see][for recent calculations corroborating this picture]{eprgv2015}.  
This implies that in not-too-old INSs,
with age $\la$ a few Myrs, the evolution of the core field can be ignored. 
This applies in particular to the M7, which have a dynamical age $\sim 0.5\
\mathrm{Myrs}$ \cite[e.g.][and references therein]{mignani13}. 
Observationally, the picture of field decay is, however, far from being
clear.  The properties of the radio-pulsar population can be reproduced
without invoking any decay of the magnetic field, or assuming a timescale so
long that it exceeds the lifetime of a normal radio pulsar
\cite[e.g.][]{rdfp01,fgk06}.  On the other hand, some studies have shown that
synthetic populations match the observed ones if the field decays on a
timescale of a few Myrs (\citealt{gonetal02}).  While these calculations
assumed a constant field decay rate, more recent population synthesis
simulations rely on a more accurate, self-consistent modelling of the field,
spin, and thermal evolution, providing results in correspondence with data
\cite[][]{petal2010, gullon14}.  Finally, \cite{ip14} demonstrated that a
brief episode of
relatively rapid field decay (characteristic timescale few $\times10^5$~yrs)
can fit well data on young and middle age radio pulsars.

We mention also that, aside from the evolution of the field itself, the
rotational properties of an INS are influenced by possible variations in
time of the direction of the dipole moment with respect to the star spin
axis (which is measured by the angle $\xi$, see Section \ref{thermal}). 
This issue has been extensively discussed in the literature, partly in
connection with magneto-rotational evolution (see \citealt{phil2014} and
references therein), partly in
connection with the precession of INSs \cite[e.g.][for a review]{link03}. 
In particular, it has been shown that the star tends to a global minimum
energy configuration which is characterized by $\xi=0$ or $\xi=\pi/2$
\cite[i.e.  either an aligned or an orthogonal rotator; e.g.][and references
therein]{was03}.  While local minimum energy states (in which the INS
precesses) can be achieved over relatively short timescales, the global
equilibrium (when no precession occurs) requires much longer, as compared to
the spin-down time.  We note that the spin-down of an INS (and its thermal
history) can be influenced by other effects, besides magneto-rotational
losses.  They include gravitational wave emission, the interaction with a
fossil disc and possibly the interaction of the rotating dipole with the
magnetized vacuum around the star \cite[the so-called quantum vacuum
friction; e.g.][]{davies05,xiong15}.  These are not addressed in the present
investigation.

Here we focus on the evolution of the crustal component of the field and
assume that the core field does not change in time.  Moreover, given that no
precession is observed in \src\footnote{Precession is indeed measurable in
the M7, and it was suggested to occur in RX J0720.4-3125 \cite[][and
references therein]{hohle12}.}, we do not consider any evolution of $\xi$.  
Magnetic field evolution in a INS crust proceeds mainly under the
influence of two processes: Ohmic dissipation and Hall cascade (see, for
example, \citealt{cumming2004}).  Ohmic dissipation can be due to electron
scattering on phonons while the temperature is high enough.  The
characteristic timescale for the field evolution at this stage is $\sim 1$
Myr, or even shorter depending on the magnetic field.  After the temperature
drops below a critical value dissipation is due to impurities, and the
timescale is significantly longer.  Note that these timescales are estimated
for the large scale magnetic field.  The Ohmic timescale is
(\citealt{cumming2004}):

\begin{equation}\label{ohm}
\tau_\mathrm{Ohm}=\frac{4 \pi \sigma d^2}{c^2}=4.4\times10^6
\left(\frac{\sigma}{10^{20} \mathrm{s}^{-1}}\right)\left(\frac{d}{1\,
\mathrm{km}}\right)^2\ \mathrm{yr}
\end{equation}
where $\sigma$ is conductivity, $d$ the typical lengthscale of the
field structure, and $c$ the speed of light. Since $d$ becomes
smaller for higher-order multipolar components, small scale fields
dissipate much faster.

The Hall cascade is a non-dissipative process (\citealt{gr1992}).
It just redistributes magnetic energy, mainly transferring it from
the larger to the smaller scales, where it is rapidly dissipated
due to Ohmic losses. The typical timescale of the Hall process can
be estimated as (\citealt{cumming2004}):

\begin{eqnarray}\label{hall}
\tau_\mathrm{Hall}&=&\frac{4\pi n_\mathrm e e d^2}{cB}\nonumber\\
&&\\
&=& 6.4 \times 10^5
\left(\frac{n_\mathrm e}{10^{36}\, \mathrm{cm}^{-3}}\right)
\left(\frac{d}{1\, \mathrm{km}}\right)^2 \left(\frac{B}{10^{14}\,
{\mathrm G}}\right)^{-1}\ \mathrm{yr},\nonumber
\end{eqnarray}
where $n_\mathrm{e}$ is the electron density, and $d$ is again the typical
lengthscale for $B$ (and so for currents and $n_\mathrm{e}$) in the crust.
Note, that $n_\mathrm{e}$ and $d$ can vary, and for NSs with a large
magnetic field $t_\mathrm{Hall}$ can be small. So, it is expected
that highly magnetized NSs undergo rapid field evolution. This is
now believed to be the case for the magnetar candidates, in which
rapid field dissipation is driven by field reconfiguration due to
Hall cascade (\citealt{v2013}).

Calculations by \cite{gc2014_2, gc2014}, under the assumption of
axial symmetry,  have shown that Hall evolution saturates after a
few $t_\mathrm{Hall}$. The magnetic field reaches some stable
configuration, and the successive evolution is driven mainly by the
relatively slow Ohmic dissipation. The stage when the Hall cascade
stops is called by these authors the {\it Hall attractor} and it
is reached in $\la 1$ Myr for magnetar-like initial fields. These
picture was confirmed by the 3D numerical simulations by
\cite{wh2015}. Due to the Hall cascade the field in the crust is moved
towards the
crust-core boundary, and dissipates there (\citealt{gc2014}). Thus, field
looks more like a core-centered field.
According to \cite{gc2014} the Hall attractor has a
well-defined property. When the field structure is stabilized, its
poloidal part mainly consists of dipole and octupole components
(with small addition of the $l=5$ multipole).

Since the surface
thermal distribution in a cooling, magnetized INS is determined by
the structure of its crustal field \cite[e.g.][see Section
\ref{thermal}]{green83, page95, gepp04, gepp06} the analysis of
INS X-ray spectral properties can provide a direct test for the
Hall attractor scenario.

\section{Populations of isolated neutron stars}
\label{pop}

\begin{figure*}
\begin{center}
\includegraphics[width=16cm]{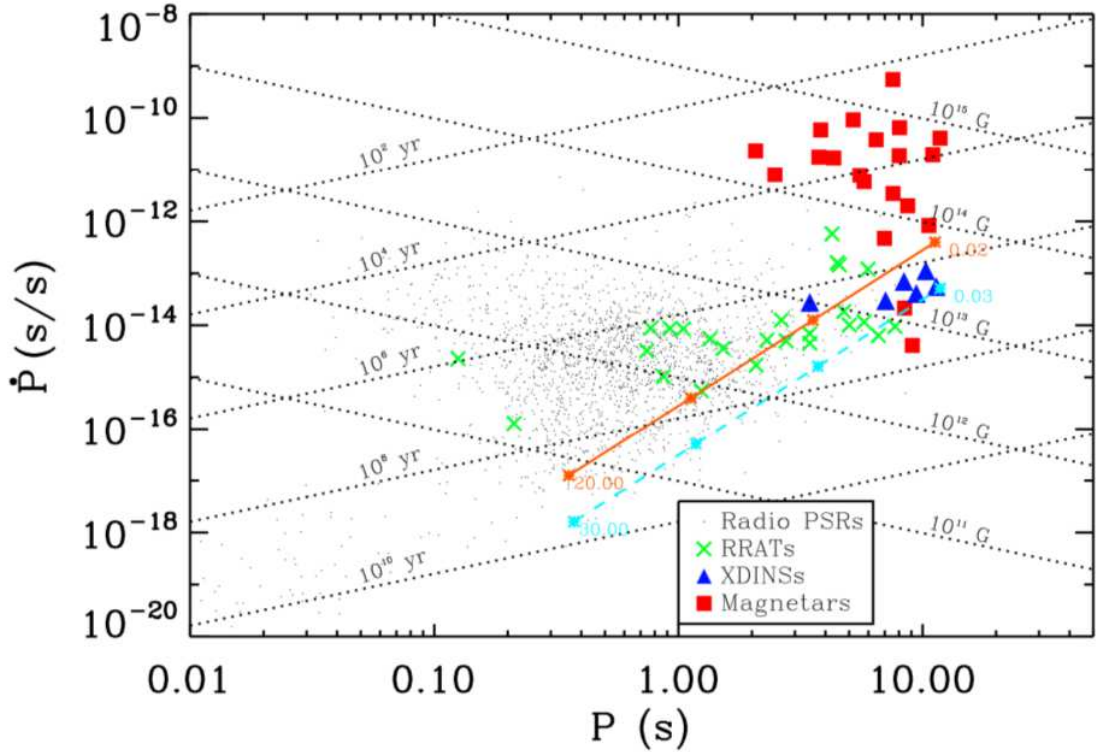}
\caption{The $P$-$\dot P$ diagram showing the different INS
classes. The full orange (dashed light blue) line corresponds to
an age equal to two  (three) Hall timescales (see text); here an
initial period $P_0=0.01\, \mathrm s$ has been assumed and the
initial magnetic field is in the range $10^{12}\, \mathrm G \leq
B_0\leq 10^{15}\, \mathrm G$. The asterisks along the two lines
mark the true age of the star in Myrs, the spacing between two
symbols corresponds to a factor of 10 decrease, moving from left
to right. } 
\label{ppdot}
\end{center}
\end{figure*}

In order to compare with observations the predictions on the field
evolution by Gourgouliatos \& Cumming, we need to select a sample
of INSs which are allegedly close to the Hall attractor stage. In
particular, these sources need to be of the right age ($\la 1$
Myr), possess initially high fields,
 and should display  thermal emission from their cooling
surface (better, not polluted by significant non-thermal
magnetospheric contribution). The magneto-rotational evolution of
an initially highly magnetized INSs can be approximately followed
assuming that during the Hall stage the magnetic field decays
exponentially with a characteristic time $\tau=10^4  (B/10^{15}\,
\mathrm G)^{-1}\ \mathrm{yr}$ (see, for example,
\citealt{aguilera08} for discussion of simple fits for the field
decay model in the presence of the Hall term). In this case the
loci of constant age in the $P$-$\dot P$ plane do not coincide
with the usual constant characteristic age lines. They are shown
in figure \ref{ppdot} for an age equal to two (orange solid line) and
three (light blue dashed line) Hall timescales, i.e. when the magnetic
field has decayed by a factor $\exp{(-2)}$ and $\exp{(-3)}$,
respectively. The asterisks along the two lines give the true age
of the star in Myr (for the initial spin period equal to 0.01 s).
Note that an INS will reach the line at different times according
to its initial magnetic field, since the Hall timescale depends on
$B_0$.

According to figure \ref{ppdot}, while some old radio pulsars can indeed
reach the Hall attractor, on average objects of this type have
too low magnetic fields, and hence too long Hall timescales, in
agreement with the results by \cite{ip2015}.
Active magnetars appear also to be far from the
attractor stage, as expected, since after the attractor is reached the
release of magnetic energy, and hence the activity,  should
decrease (see, however, \citealt{wh2015}, who claim that old
magnetars can still produce bursts, even after reaching the
attractor stage).

Quite interestingly, figure \ref{ppdot} shows that close-by,
cooling isolated NSs, the XDINSs or M7, lie close to the Hall
attractor stage, as quite naturally follows if these sources are
descendants of magnetars (so that their initial field was huge), with typical ages $\sim 0.5$~Myr
(\citealt{petal2010}). This is very fortunate, as the M7 are
purely thermal sources, and their surface emission is potentially
sensitive to the magnetic field structure in the crust.
Indirectly, the proximity of the M7 to the attractor may be also
supported by their low activity. In the next subsection we briefly
summarize the properties of X-ray emission from the M7.

\subsection{X-ray emission from the Magnificent Seven}
\label{7x}


The M7 are isolated neutron stars characterized by stable thermal
emission with temperatures $\sim 50$-$100$~eV\footnote{See the
on-line catalogue of thermally emitting NSs at
http://neutronstarcooling.info \cite[][]{v2013}.}. Typically, their
spectra can be fitted by one or two blackbody components with the
addition of a broad absorption feature at few hundred eVs. The pulsed fraction is
usually low ($\la 20$\%) and the luminosities are in the range
$L\sim10^{31}$--$10^{32}$~erg~s$^{-1}$. A recent summary of the
main properties of the M7 can be found in \citet[see also \citealt{turolla09}]{pires2014}.

The spin-down measure of the dipole field in the M7 gives values somehow in excess 
of those typical of radio pulsars and not too far from those of the magnetars, i.e. $10^{13}$--$10^{14}$ G \cite[e.g.][]{turolla09}\footnote{These values are in agreement with those
obtained from the energy of the absorption features, assuming that the latter are produced by proton cyclotron resonance
or by bound-bound transitions in low-Z elements.}. This led to the suggestion that the M7 could be elderly
magnetars, kept hotter than normal INSs of the comparable age by field decay \cite[][]{ponsetal07}. Indeed, detailed population synthesis calculations
confirmed that the general properties of the M7 follow naturally in the framework of a decaying magnetic
field \cite[][]{petal2010}.

\begin{figure}
\begin{center}
\includegraphics[width=9cm]{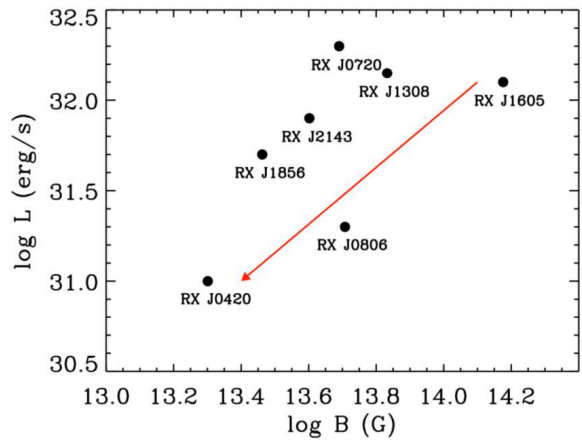}
\caption{The placement of the M7 in $B$--$L$ plane (data taken
from http://www.neutronstarcooling.info/). The red arrow indicates
the likely evolutionary path of a NS during the Hall phase.}
\label{lumB}
\end{center}
\end{figure}

Different sources among the M7 can be at different evolutionary
stages, and so they can be closer or farther from the attractor
stage. This may translate into different levels of activity,
since, according to \cite{gc2014} sources close to the attractor
should appear less active. Actually, none of the M7 exhibits
magnetar-like activity. Still, they display somehow different
characteristics, and this can be used to gauge the proximity to
the attractor (see Fig. \ref{lumB}). In particular, NSs at (or very near) the attractor
should have lower pulsed fraction and lower magnetic field than
other sources of the same class, together with lower temperatures
and luminosities, since it is expected that, in the case of the
M7, temperatures and luminosities are slightly enhanced due to
field decay (\citealt{petal2010}). Following this line, RX
J0720.4-3125 may provide a good case for an M7 source which is still
far from the Hall attractor. In fact, it has the largest
luminosity among the Seven, its pulsed fraction is $>10$\%, and it
exhibits long-term spectral variability \cite[e.g.][and references
therein]{vk07,hohle12}. On the other hand, RX J0420.0-5022 and RX
J1856.5-3754 may be representatives of sources close to the Hall
equilibrium: they have low blackbody temperatures ($\sim 50$--$60$
eV), and hence luminosities, and their dipolar B-field, $\sim
10^{13}\, \mathrm G$, is the weakest among the Seven.

At variance with RX J0420.0-5022, RX J1856.5-3754 is the prototype
and brighter member of the class, and its spectral properties are
very well-characterized. The pulsed fraction is the lowest, $\sim
1$\%, no variability was detected so far and the X-ray spectrum is
well fitted by two blackbody components with $kT^\infty_1 \sim
61$--$62$~eV, $R^\infty_1\sim 4.5$--$5$~km and $kT^\infty_2 \sim
39$~eV, $R^\infty_2\sim 11$--$16$~km \cite[][]{sart2012}; all the
previous quantities are referred to an observer at infinity. For
these reasons in the following we focus on RX J1856.5-3754 (RX
J1856 hereafter), and try to assess if its spectral properties are
indeed compatible with the crustal distribution of the magnetic
field predicted by the Hall attractor configuration.

\section{The model}
\label{thermal}

In order to compute synthetic spectra from the cooling surface of
an isolated NS we follow the same approach presented in \citet[see
also \citealt{page95,taverna15}]{zane06}. The star surface is
divided in a number of patches by means of an equally-spaced grid
in $\mu\equiv\cos\Theta$ and $\Phi$ ($0\leq \mu\leq 1$, $0\leq
\Phi\leq 2\pi$), where $\Theta$ and $\Phi$ are the co-latitude and
the azimuth relative to the line-of-sight (LOS). Relativistic
ray-bending (the vacuum Schwarzschild solution is assumed
throughout) is accounted for by introducing the angle $\bar\Theta$,
\begin{equation}
\bar\Theta =
\int_0^{1/2}\frac{dv\, \sin\Theta}{
\left[(1-x)/4-(1-2vx)v^2 \sin\Theta^2 \right]^{1/2}}\,,
\end{equation}
so that the monochromatic flux emitted by the portion of the surface which is into view at a given spin phase $\gamma$ is given by
\begin{equation}
\label{flux}
F_\nu(\gamma)=(1-x)
\frac{R_{\mathrm{NS}}^2}{D^2}
\int_0^{2\pi}d\Phi\int_0^1 I_\nu(\mathbf k,\theta,\phi)
du^2\,;
\end{equation}
here $u=\sin\bar\Theta$, $D$ is the source distance, $x=R_\mathrm
s/R_\mathrm{NS}$, $R_\mathrm s=2GM_\mathrm{NS}/c^2$ and
$M_\mathrm{NS}$, $R_\mathrm{NS}$ are the star mass and radius. In
equation (\ref{flux}), $I_\nu$ is the specific intensity which in
general depends on the photon direction $\mathbf k$, and on the
position of the emitting point on the surface, expressed here through
the magnetic co-latitude $\theta$, and azimuth $\phi$. The
components of $\mathbf k$, $\theta$ and $\phi$ can, in turn, be
expressed in terms of $\Theta$, $\Phi$, $\gamma$ and the two
geometrical angles $\chi$ and $\xi$ which give, respectively, the
inclination of the LOS and of the magnetic axis with respect to
the star spin axis.

In the following we restrict to magnetic configurations described
by the superposition of multipoles up to $l=5$ (see section
\ref{hall}); we specify the strength of the dipole at the magnetic
pole, $B_\mathrm p$, and the polar ratio of the $l$-multipole to
the dipole, $\rho_\mathrm l$. General-relativistic corrections to
the $B$-field were included through the functions $f(x)$ and
$g(x)$, for the $r$- and $\theta$-component of the field,
respectively \cite[][]{muslimov86}\footnote{Note that for higher
order multipoles, the analytical expression for the hypergeometric
functions, which give $f$ and $g$, becomes numerically unstable
even for $x\sim 0.3$. For this reason we used the sum of the
(truncated) hypergeometric series instead.}. For magnetic fields
$\ga 10^{11}\ \mathrm G$ electron thermal conduction is
essentially along the magnetic field lines and the surface
temperature distribution is given by
\begin{equation} \label{cctemp}
T_\mathrm{s}\simeq T_\mathrm p\vert\cos\theta_\mathrm
B\vert^{1/2}\,,
\end{equation}
where $T_\mathrm p$ is the polar value of the temperature and
$\theta_\mathrm B$ is the angle between $\mathbf B$ and the
surface normal \cite[e.g.][see also
\citealt{potekhin15}]{green83,page95}.

No definite physical model for the surface emission from the M7
has been put forward as yet \cite[see e.g.][for a
discussion]{potekhin14}. It has been suggested that in these
sources the surface layers are in a condensed state, owing to a
phase transition driven by the low surface temperature and
relatively high magnetic field \cite[][see also
\citealt{turolla09},
\citealt{potekhin14}]{laisalp97,burwitz03,turolla04,medinlai07}.
In particular, \cite{ho07} used the condensed surface emission
model to explain the multiwavelength spectral energy distribution
of RX J1856, with the addition of a thin, magnetized, H atmosphere
on top of the condensate to reproduce the optical spectrum. In the
next section we present results for the X-ray spectrum of RX J1856
for magnetic configurations typical of the Hall attractor for both
(isotropic) blackbody emission at the local temperature $T$ and a
condensed surface. The specific intensity in the former case is
simply
\begin{equation}
I_{\nu,{\mathrm
{BB}}}=B_\nu(T)=\frac{2h}{c^2}\frac{\nu^3}{\exp(h\nu/kT)-1}
\end{equation}
while in the latter it can be expressed as
\begin{equation}
I_{\nu,\mathrm C}=j_{\nu}(B,\mathbf
k,\theta_\mathrm{Bk})B_\nu (T)\,,
\end{equation}
where $j_\nu$ is the emissivity of the condensed phase and
$\theta_\mathrm{Bk}$ is the angle between the photon direction and
the magnetic field; the analytical approximations by
\cite{potekhin12} were used to compute $j_\nu$ \cite[see also][for
more details]{gonzalez16}. Because of the present uncertainties in
modelling the dielectric tensor of the condensed phase, here we
consider both the two limiting cases in which ions are either
treated as ``free'' or ``fixed'' \cite[see e.g.][for more
details]{turolla04,potekhin12}.

\section{Numerical results}
\label{results}

The spectral properties of RX J1856 were simulated assuming a NS
with mass $M_\mathrm{NS}=1.4$ M$_\odot$ and radius
$R_\mathrm{NS}=12$ km; we also assumed a polar temperature at the
surface $T_\mathrm{p}\simeq 75$ eV, compatible with the estimated
temperature of the hotter blackbody at infinity by \cite[see \S
\ref{7x}] {sart2012}\footnote{The gravitational redshift factor
for the values of mass and radius used here is $1+z\sim 1.24$.}.
The temperature profile given by equation (\ref{cctemp}) was
slightly modified, by truncating it at $T_\mathrm{e}=5$ eV in
order to avoid a vanishing temperature at the magnetic equator
\cite[see e.g.][]{taverna15,gonzalez16}. Following \citet{gc2014},
we mimicked the Hall attractor stage adopting two different
magnetic field topologies: a combination of dipolar and octupolar
components, with opposite polarity and polar ratio $\rho_3=0.6$,
and a similar configuration, with the addition of an $\ell=5$
component, characterized by the polar ratios $\rho_3=0.6$ and
$\rho_5=0.3$. Hereafter we will refer to these two models as model
1 and model 2, respectively. According to the measured values of
$P$ and $\dot{P}$, we set the magnetic field intensity for the
dipolar component at the poles at \mbox{$B_\mathrm{p}\simeq
10^{13}$ G} \cite[][]{kk08}.

\begin{figure*}
\begin{center}
\includegraphics[width=1.\textwidth]{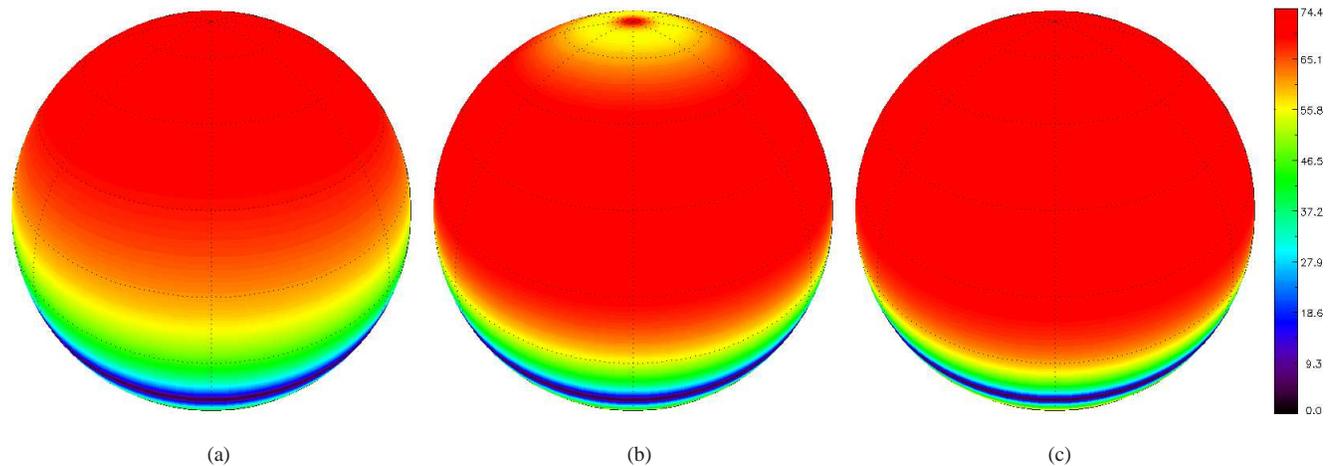}
\setlength{\tabcolsep}{75pt}
\begin{tabular}{cccc}
\ & \ & \ & \ \\
(a) & (b) & (c) & \ \\
\ & \ & \ & \ \\
\end{tabular}
\caption{Temperature maps for the three different magnetic field
topologies discussed in the text: pure dipole (left), dipole plus
octupole (model 1, center) and dipole plus octupole plus $\ell=5$
component (model 2, right).} \label{fig:tempmaps}
\end{center}
\end{figure*}

The resulting maps for the surface temperature are shown in Fig.
\ref{fig:tempmaps}b for model 1 and Fig. \ref{fig:tempmaps}c for
model 2; the map for a purely dipolar magnetic field is also shown
for comparison. Model 1 and 2 only differ for a small region
around the magnetic pole: here the surface temperature is $\sim
25\%$ lower with respect to $T_\mathrm{p}$ for model 1, while, at
the same magnetic colatitudes, it is nearly constant (and equal to
the polar value) for model 2. On the other hand, comparing the
first two maps with that for the purely dipolar field, it turns
out that the addition of higher-order multipoles generally
increases the surface temperature, making its distribution more
uniform. This implies that the pulsed fraction will likely
decrease as the star approaches the Hall attractor.

\begin{figure*}
\begin{center}
\includegraphics[width=1.\textwidth]{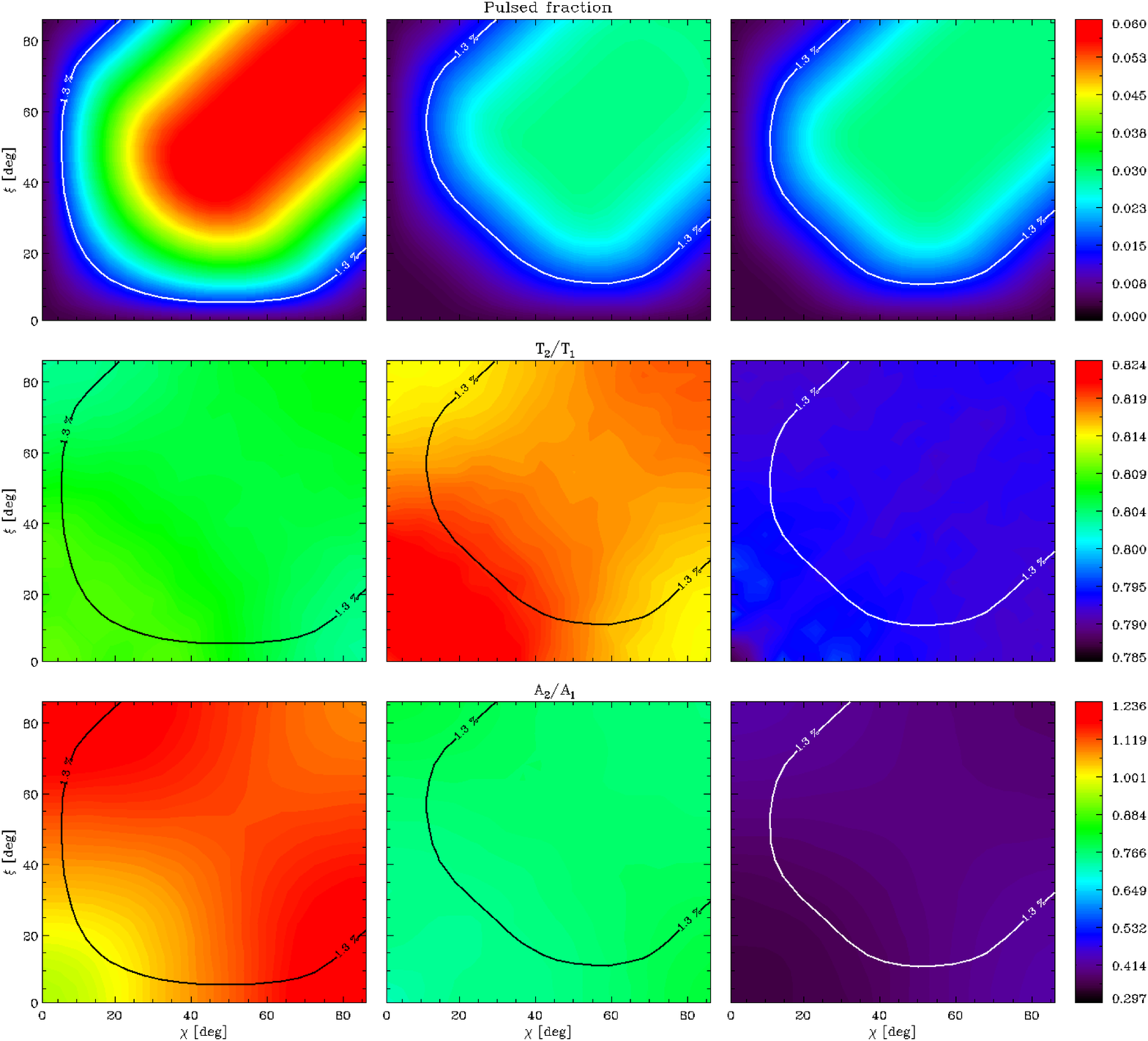}
\caption{Contour plots for the pulsed fraction (top row), the
ratios of the temperatures (middle row) and emitting areas (bottom
row) of the two blackbody components which fit the phase-averaged
spectra, plotted as functions of the viewing angles $\chi$ and
$\xi$ for a pure dipole (left column), model 1 (middle column) and
model 2 (right column); here the pulsed fraction is computed in
the 0.1-1.6 keV energy range and local emission is blackbody. The
curve corresponding to the observed pulsed fraction for RX J1856
($\sim 1.3\%$) is also shown.} \label{fig:fits}
\end{center}
\end{figure*}

We then simulated the emission, first considering the case of
surface blackbody radiation with temperature $T=\max(T_\mathrm{s},
T_\mathrm{e})$ (see equation \ref{cctemp}). The top row of Figure
\ref{fig:fits} shows the pulsed fractions, in the 0.1--1.6 keV
energy range, for the pure dipole (left), model 1 (center) and
model 2 (right), plotted as function of the angles $\chi$ and
$\xi$. The decrease of the pulsed fraction as higher-order
multipoles are added to the dipole is apparent from Figure
\ref{fig:fits}. The loci in the $(\chi,\xi)$ plane that correspond
to the observed value of the pulsed fraction for RX J1856 ($\sim
1.3\%$, see \S \ref{7x}), are also drawn to show the geometrical
configurations which correspond to the observational data.

In order to check if the spectral parameters derived by
\citet{sart2012} are indeed reproduced, we calculated the phase-averaged spectra in the
0.1--2 keV energy range for each pair of angles $\chi$ and $\xi$
and fitted them with two blackbody components at temperature $T_1$
and $T_2$
\begin{equation} \label{fittingfunc}
F = A_1E^3\left(\frac{1}{\exp{(E/kT_1)}-1}+\frac{A_2/A_1}{\exp{(E/kT_2)}-1}\right)\,,
\end{equation}
where the normalization $A_1$ ($A_2$) depends only on the emitting
area. The middle and bottom rows of Figure \ref{fig:fits} show,
again as a function of $\chi$ and $\xi$, the behavior of the
ratios $T_2/T_1$ and $A_2/A_1$, respectively, for the same three
magnetic configurations described above. Moving from the dipole
towards the Hall attractor stage, the ratio $T_2/T_1$ increases
slightly, being anyway close to $\sim 0.7$, not that far from the
value ($\sim 0.65$) derived by \citet{sart2012}. On the contrary,
the ratio $A_2/A_1$ generally decreases. In particular, only in
the case of the dipole, and for some viewing geometries (small
$\chi$ -- large $\xi$ and vice versa), the emitting area of the
colder component is larger than the hotter one, with $A_2\sim
1.2A_1$. Adding higher-order multipole components, instead, $A_2$
becomes systematically smaller than $A_1$, resulting in $A_2\sim
2/3\,A_1$ for model 1 and $A_2\sim 1/3\,A_1$ for model 2. This
behavior contrasts with the observations, which give an emitting
area ratio in the range $\sim 6$--$10$.

\begin{figure}
\begin{center}
\includegraphics[width=0.47\textwidth]{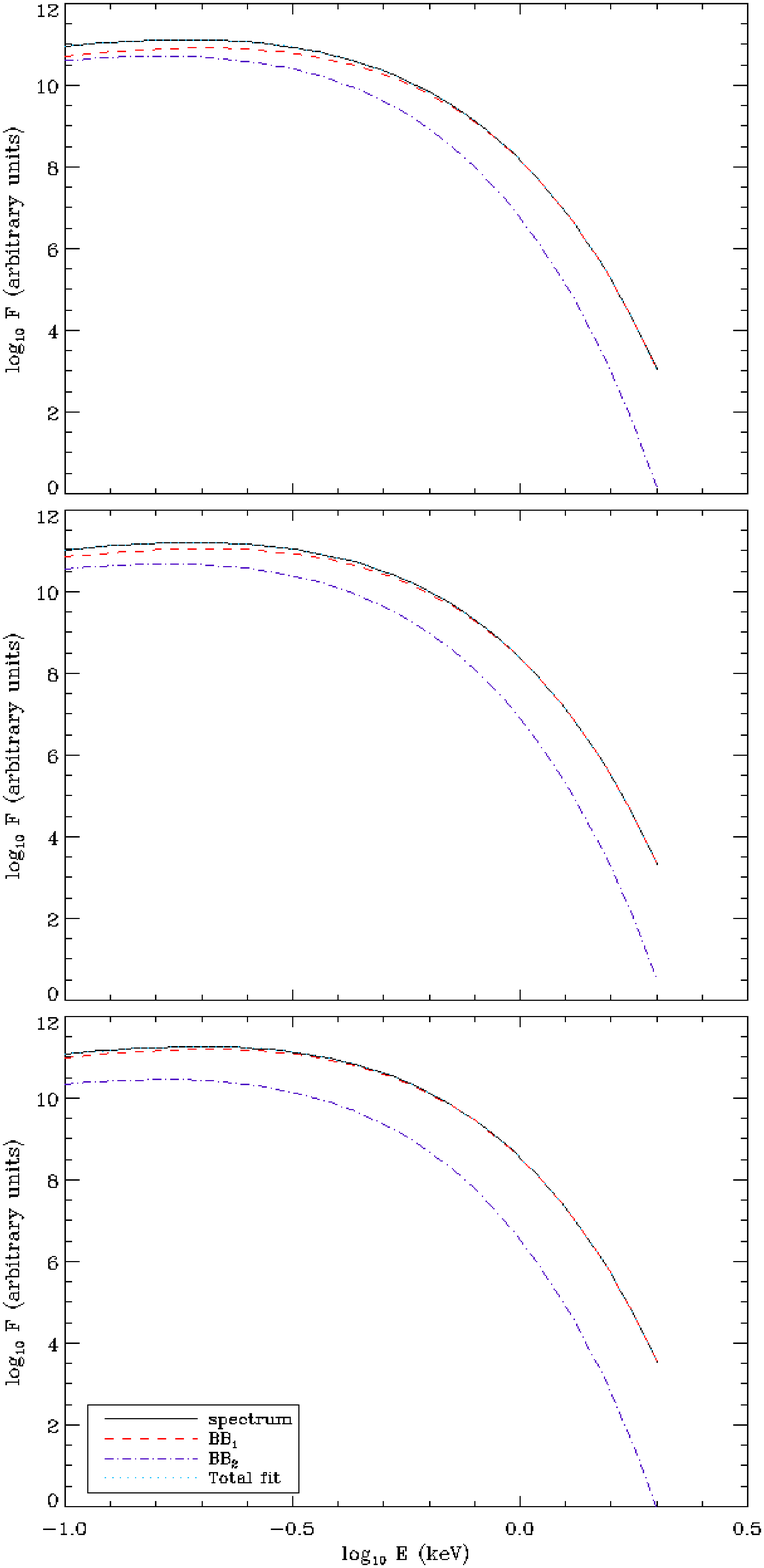}
\caption{Phase-averaged spectra in the 0.1--1.6 keV energy range
(black solid lines) in the cases of the  dipole (top panel), model
1 (middle panel) and model 2 (bottom panel), for $\xi\simeq
80^\circ$ and $\chi\simeq 15^\circ$, $20^\circ$ and $25^\circ$,
respectively (see Figure \ref{fig:fits}). The two blackbody
components (red dashed line and blue dash-dotted line) used to fit
the spectra, as well as the best-fitting curve (green dotted line)
are also shown. In all the three cases local emission is  an
isotropic blackbody (see text for details).} \label{fig:spec1D}
\end{center}
\end{figure}

As an example, Figure \ref{fig:spec1D} shows the spectra, as
observed at infinity, for the dipole (top), model 1 (center) and
model 2 (bottom), for values of the angles $\chi$ and $\xi$ chosen
in such a way that the  pulsed fraction is compatible with the
observed value (see Figure \ref{fig:fits}). The individual blackbody
components used to fit the spectra are also shown; the fit results
are summarized in Table \ref{fittable}. As discussed above, it is
clear that there is no way to reproduce the observed emitting area
ratio $A_2/A_1$ with the magnetic field topologies we used to
approximate the Hall attractor stage, although in all the three
cases spectra are well fitted with two blackbody curves with
temperatures compatible with those derived by \citet{sart2012}.

\begin{table}
\begin{center}
\vspace*{-\baselineskip}
\begin{tabular}{lccccc}
\hline
\ & $\chi$ & $\xi$ & $T_1$ (eV) & $T_2$  (eV) & $A_2/A_1$ \\
\hline
Pure dipole & $15^\circ$ & $80^\circ$ & 72.0 & 57.8 & 1.27 \\
\hline
Model 1 & $20^\circ$ & $80^\circ$ & 73.0 & 59.4 & 0.76 \\
\hline
Model 2 & $25^\circ$ & $80^\circ$ & 73.5 & 58.1 & 0.36 \\
\hline
\end{tabular}
\caption{Results of the fits performed for the spectra shown in
Figure \ref{fig:spec1D}. The values of temperature are  at the
star surface.} \label{fittable}
\end{center}
\end{table}

\begin{figure}
\begin{center}
\includegraphics[width=0.47\textwidth]{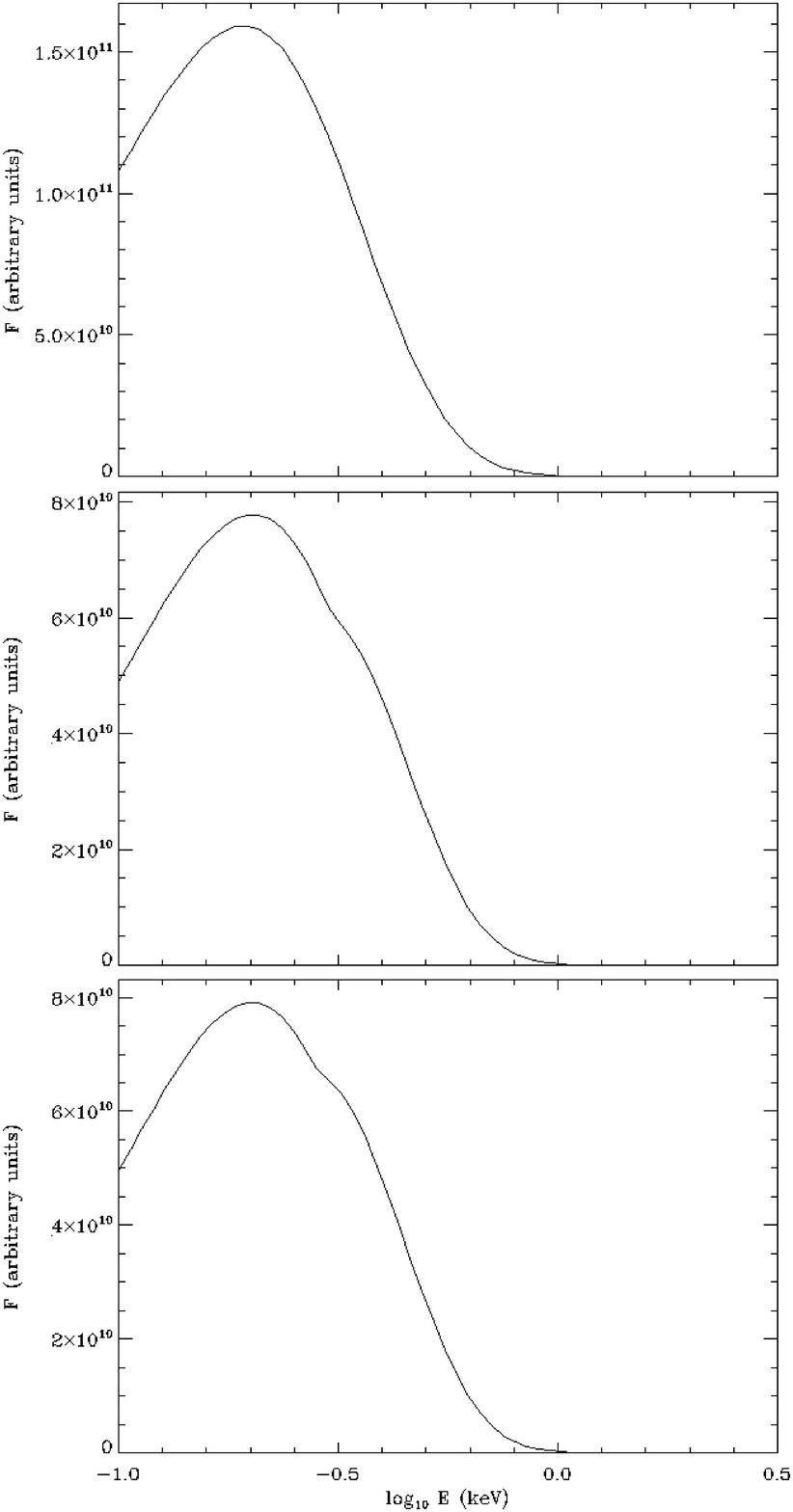}
\caption{Phase-averaged spectra in the 0.1--1.6 keV energy range in
the case of model 1 for different surface emission models:
blackbody (top panel) and condensed surface (free-ions, middle
panel; fixed-ions, bottom panel). The viewing angles are chosen in
such a way that the corresponding pulsed fractions are compatible
with the observed value ($\sim 1.3\%$), i.e. $\chi\simeq
20^\circ$, $\xi\simeq 80^\circ$ for the blackbody  and $\chi\simeq
25^\circ$, $\xi\simeq 5^\circ$ for the condensed surface.}
\label{fig:bbfrfx}
\end{center}
\end{figure}

Finally, we performed the same simulations assuming that emission
is from a condensed surface, using the analytical approximations
by \citet{potekhin12}, both in the free-ions and fixed-ions
limits. We found that, for all the magnetic field configurations
considered, a fit with two blackbody components (as given by
equation \ref{fittingfunc}) fails match the observations,
with a ratio $A_2/A_1$ smaller than $1 \%$. In particular, Figure
\ref{fig:bbfrfx} clearly shows that, for the case of model 1, the
shape of the spectra obtained for the condensed surface emission
model (middle and bottom panels) requires the presence of at least
another component in order to explain the spectral hardening that
is present at energies around 0.4--0.5 keV, at variance with what
happens for the isotropic blackbody emission (top panel), where
the spectral enhancement is not present. Similar results have been
found in the cases of model 2 and purely dipolar field.

In summary, our results indicate that for the field configuration
predicted by the Hall attractor, the area corresponding to the
hotter temperature is larger than that at the cooler temperature,
contrary to what is observed in RX J1856.

\section{Discussion}
\label{disc}

It has been recently suggested that the magnetic field evolution
in isolated neutron stars proceeds through a phase dominated by
the Hall cascade until a Hall equilibrium stage (the Hall
attractor) is reached \cite[][]{gc2014_2,gc2014,wh2015}. The Hall
attractor is characterized by a well defined structure of the
crustal magnetic field and, in turns, of the surface temperature
distribution. The latter can, in principle, be probed through
X-ray observations of thermally emitting INSs, and, in this
respect, the ``Magnificent Seven'' (M7) provide an optimal target.
In this paper we have compared the predictions of the Hall
attractor scenario with the observed spectral properties of the
brightest of the M7, RX J1856.5-3754. Our main conclusion is
that the surface temperature distribution produced by the Hall
attractor magnetic field configuration fails to explain the X-ray
spectrum of RX J1856.5-3754, as derived from {\em XMM-Newton}
observations by \cite{sart2012}. In particular, while the
temperatures of the two blackbody components which best-fit the
data are broadly reproduced, the emitting areas are not.

The nature of the surface emission from cooling neutron stars is
still not completely understood. For this reason we considered
both isotropic blackbody emission and emission from a condensed
phase, both at the local temperature. Agreement with observations
was not found in either case, irrespectively of the geometry of
the source (inclination of the line-of-sight and of the magnetic
axis with respect to the rotation axis). Although it can not be
excluded that other emission models can provide a better agreement
with data, we deem this unlikely. 

General-relativistic ray-bending was properly accounted for in our
calculations. Numerical results were obtained for reasonable
values of the star mass and radius, $1.4M_\odot$ and $12$ km,
respectively, and hence for the $M/R$ ratio. However, since no
estimate for these two quantities is available for RX
J1856.5-3754, one may wonder if other choices for $M_\mathrm{NS}$
and $R_\mathrm{NS}$ could change our conclusions. While we did not
attempt a systematic exploration, results obtained for other
values, e.g. $M_\mathrm{NS}=1.2M_\odot$ and $R_\mathrm{NS}=15\,
\mathrm{km}$, are in qualitative agreement with  previous ones,
with only some marginal quantitative differences.

\subsection{Comparison between RX J1856 and RX J0720}


It is interesting to compare properties of the two most studied sources
among the M7: RX J1856 and RX J0720.
In particular, which one is older, and what can we say about their
initial parameters and evolution.

For both objects age determinations are not very precise.
The characteristic age, $\tau_\mathrm{ch}=P/2\dot P$, is slightly larger for RX
J1856: $\log \tau_\mathrm{ch}=6.58$ vs. 6.28 (\citealt{pires2014}). But the
kinematic age is larger for RX J0720: $\log \tau_\mathrm{kin}=5.93$ vs. 5.62
(\citealt{pires2014}). Note, that RX J1856 has slightly smaller spin period
and smaller $\dot P$ (and so, smaller magnetic field $B\propto \sqrt{P\dot P}$).

In our considerations related to the magneto-rotational evolution we noted
that RX J1856 is less active (it has smaller pulse fraction, luminosity,
and temperature, it does not demonstrate strong variability, etc.).
In Figs. \ref{ppdot}, \ref{lumB} RX J1856 is situated closer to the Hall
attractor.

Taking all together, we have to conclude that RX J1856 is younger (smaller
$\tau_\mathrm{kin}$), but more evolved. It is possible if its magnetic field
was initally larger, and then the source evolved faster. However, in
standard (simplified) models of evolution with field decay,
if two NSs have similar (small) initial periods,
but significantly different magnetic fields, then the less magnetized cannot
attain longer spin period and, at the same time, higher $\dot P$. Then, some
complications to explain the data on RX J1856 and RX J0720 are necessary.
To give a visual impression, we can say that evolutionary tracks
of RX J1856 and RX J0720 might cross on the $P-\dot P$ diagram.
Potentially, this can be related to some non-trivial field evolution, so
that RX J1856 experienced a period of very rapid field decay in its past.

\subsection{How to find INSs at the Hall attractor}


It seems that even RX J1856 is not at the stage of the attractor, yet (if
predictions are correct). Then it is necessary to find more evolved
relatives of the M7.

If the Hall cascade saturates, then the field evolution is
governed only by the Ohmic processes. As a NS cools down
scattering of electrons on phonons becomes less important. There
is a critical value of the temperature, $T_\mathrm{U}$ (see, for
example, \citealt{cumming2004} and references therein), below
which the Ohmic dissipation is determined only by impurities.
Then, the field decays very slowly, and so the crust is not
additionally heated. It seems that mostly NSs at the Hall
attractor might be relatively cool sources. It would be very
difficult to detect them even with the {\it{eROSITA}}
(\citealt{erosita}), as this instrument is fitted for larger
temperatures.


Still, taking into account that the solar proximity is overabundant in NSs
with ages $\sim 10^6$~yrs due to the Gould Belt (\citealt{petal2003}),
we can expect that dozens of INSs with $T\sim$~a few tens of eV are floating
in space around us.

Another option is to look for thermal emission of RRaTs and pulsars which
are potentially close to the Hall attractor (see Fig. \ref{ppdot}) and have
ages $\la$~Myr. For near-by
sources ($\la 1$--$2$~kpc) with long exposures it is possible
to detect thermal emission in X-rays.

It seems that observations of thermal emission is one of the best way to
probe the existence of the Hall attractor and test its predicted properties.

\section{Conclusion}
\label{concl}

 In this paper we tried to verify predictions for the magnetic field
evolution in NSs made by \cite{gc2014}. Namely, we probed existence of
the Hall attractor using data on surface thermal emission of the source RX
J1856.5-3754. As modelling of the magnetic field structure performed by
\cite{gc2014} predicted relative contribution of different multiipoles, we
were able to calculate the expected spectral characteristics of such a NS.
We found that, contrary to observations of RX J1856, at the stage of Hall
attractor in the two blackbody approximation,
 the area corresponding to higher temperature is larger than the
area related to lower temperature. Thus, we conclude that in the case of RX
J1856 (and, most probaly, also in case of most or all other M7 sources) the
stage of the Hall attractor is not reached, or the field structure at this
stage is different from the tested predictions.

\section*{Acknowledgments}

SBP is grateful to the Department of Physics and Astronomy of the University of
Padova for financial support and hospitality during a visit when
part of this work was carried out. The work of SBP was also
partially supported by the Russian Foundation for Basic Research, project
17-02-00360. SBP thanks Andrei Igoshev for several discussions.
The work of RT is partially supported by INAF through a PRIN
grant.

\bibliography{hall_m7_r5}
\bibliographystyle{mn2e}

\end{document}